# Wind speed PDF classification using Dirichlet mixtures

Rudy CALIF[1], Richard EMILION[2], Ted SOUBDHAN[1] and Ruddy BLONBOU[1]
[1]GRER (Groupe de Recherche sur les Energies Renouvelables), Université des Antilles et de la Guyane,
France
[2]MAPMO (Mathématiques et Applications Physique Mathématique d'Orléans), UMR CNRS 6628
Université d'Orléans, France.

## Abstract:

Wind energy production is very sensitive to instantaneous wind speed fluctuations. Thus rapid variation of wind speed due to changes in the local meteorological conditions can lead to electrical power variations of the order of the nominal power output. In small grid as they exist on islands (French West Indies) such fluctuations can cause instabilities in case of intermediate power shortages. To palliate these difficulties, it is essential to identify and characterize the wind speed distribution. This allows anticipating the eventuality of power shortage or power surge. Therefore, it is of interest to categorize wind speed fluctuations into distinct classes and to estimate the probability of a distribution to belong to a class. This paper presents a method for classifying wind speed histograms by estimating a finite mixture of Dirichlet distributions. The SAEM algorithm that we use provides a fine distinction between wind speed distribution classes. It's a new nonparametric method for wind speed classification. However, we show that the wind speed distribution in each class correspond to a specific Gram- Charlier densities.

# 1. Introduction

Increasing the wind energy contribution to electrical network requires improving the tools needed to forecast the electrical power produced by wind farms, in order to proportion the network lines [16], [19]. To reach this objective, numerous studies dedicated to statistical and dynamical properties of wind velocity were developed during the last decade. To our knowledge, the results already presented in the literature only concern time scales larger than 10 minutes and more often one hour. Indeed, there has been considerable effort and a variety of statistical techniques applied to wind energy production, on time scales larger than 1 hour. Giebel in [8] reviews the major categories of forecasting models, including persistence models, neural networks, the RisØ model, the autoregressive time series model and the others. Moreover, these models are efficient on time scales ranging from 10 minutes to 1 hour [7].

However, wind speed fluctuations on time scales which are smaller than ten minutes can lead to electrical power variations of the order of the nominal power output. Indeed, rapid changes in the local meteorological condition as observed in tropical climate can provoke large variations of wind speed. Consequently, the electric grid security can be jeopardized due to these fluctuations. This is particularly the case of island networks as in the Guadeloupean archipelago (French West Indies), where the installed 20MW wind power already represents 5% of the instantaneous electrical consumption. Therefore, when wind energy becomes a significant part of the electricity networks. This percentage should reach 13% by 2010. To manage and control the electrical network and the alternative power sources, it's necessary to improve the identification of these small time scales variations, in order to anticipate the eventuality of power shortage. A first step towards the development of forecasting tools is the identification and the characterization of the wind speed density, on times scales smaller than 10 minutes. Here we develop a classification method of the different meteorological events encountered, over 10 minute periods, during the whole measurement duration. Furthermore, this study will highlight to the statistical moments of frequent and marginal wind regimes under tropical climate. The histogram classification method that we have chosen is the estimation of a mixture of Dirichlet distributions as done in [4], [18]. An overview of this method is as follows. We use 1 million of sequences of wind speed on sliding windows of 10 minutes size, obtained from a six month measurement campaign. We first convert these sequences into histograms built from a fixed partition of 12 bins on the range interval of wind speed. Each histogram is then equivalent to a probability vector of 12 nonnegative components which sum up to one. On the set of such vectors, we can put the interesting well-known Dirichlet distribution. More precisely, we will consider the 1 million histograms as a sample from a finite mixture of Dirichlet distributions that we have to estimate. Each component of this mixture will be the distribution of a class of histograms. Therefore the estimation will provide the classes that we were looking for and the probability that a given sequence belongs to each class. This method is interesting at least for two reasons. First, histograms capture the entire range of meteorological events and all statistics of wind speed sequences (e.g., all the moments and not just the first and the second moment as it is usually done). Secondly, this method is clearly nonparametric since no hypothesis is made on the shape of wind speed distributions. It seems that such an approach has never been used on wind speed data.

The paper is organized as follows. Section 2 concerns the experimental set-up of wind speed measurement. We present our motivation and our method for creating empirical histograms from wind speed measurements in section 3. In section 4, we present our model and the related background material. In section 5, we apply our model to wind speed measurement when either two or three classes are used, and we propose a mathematical function for modeling the experimental wind speed distribution. In section 6, we present an analysis of the sequence of classes.

## 2. Experimental set-up

The wind speed is measured at the wind energy production site of Petit Canal in Guadeloupe. This 10 MW production site, managed by the Vergnet Caraïbe Company, is positioned at approximately 60 m (197 ft) above sea level, at the top of a sea cliff.

<table>
<tr><th></th><th>A100L2R</th><th>W200P</th></tr>
<tr><td>Size</td><td>height=200mm<br>diameter=55mm<br>weight=350g</td><td>height=270mm<br>diameter=56mm<br>weight = 350g</td></tr>
<tr><td>Supply Voltage:</td><td>12 V (6½V to 28V)</td><td>5 V (20V max.)</td></tr>
<tr><td>Materials:</td><td colspan="2">Anodized aluminium, stainless steels and ABS plastics for all exposed parts</td></tr>
<tr><td>Range of Operation:</td><td>Threshold: 0.15 m/s<br>starting speed: 0.2 m/s<br>stopping speed: 0.1 m/s)<br>Max. wind speed: (75m/s)</td><td>Max. Speed: >75ms; range: 360° mechanical angle<br>Accuracy: ±2° obtainable in steady winds over 5 m/s.<br>(3.5°gap at North)</td></tr>
<tr><td>Analogue Output:</td><td>Calibration: 0 to 2.500 V DC for 0 to 75 m/s (32,4 mv per m/s).</td><td rowspan="2">0 to 5 V for 0° to 360°</td></tr>
<tr><td>Response Time:</td><td>150ms first order lag typical</td></tr>
</table>

**Table 1: Anemometer and wind vane specifications.**

The wind speed and direction were measured simultaneously, in a horizontal plane, with a three-cup anemometer (model A100L2 from Vector Instruments) and a wind vane (model W200p from Vector Instruments). Both were mounted on a 40 m (131 ft) tall mast erected 20m (66 ft) from the cliff edge, at 38 m (125 ft) from the ground. The response time of the anemometer is 0.15 s. This remains compatible with a sampling rate of 1 Hertz for the sake of a statistical analysis of the wind speed variations. Table 1 gives the specifications of both the anemometer and the wind vane. The measured data are downloaded to a PC connected to the RS232 port of a Campbell Scientific CR23X data logger. This data acquisition system was set-up to operate continuously and the PC can be administrated via a phone line, which allows a remote control of the data acquisition operation.

## 3. PDF – based classification

In this section we describe the data used for classification, how we built the wind speed histograms and our motivation for doing so. The data used in this paper comes from a six month measurement campaign of the wind speed at the wind energy production site of Petit-Canal in Guadeloupe (French West Indies). The measurements were carried out from December 5$^{th}$ 2003 until March 31$^{st}$ 2004 i.e. during the trade wind season. In the purpose of to identify and classify the wind speed density, on time scales ranging from 1 second to 600 seconds, the first step consists in splitting into the whole set of measurements as follows: we construct the 1,000,118 wind speed time series.

### 3.1 PDF

Thus $X_i$ gives an empirical histogram for wind speed sequence $i$ over the set of bins $B$. In Figure 2 we give a dummy example with 4 sequences and 4 bins to help clarify our notation. Each value represents a sample value for $X_{il}$. Since we have $\sum_{l=1}^{L} X_{il} = 1$, we indeed have a proper histogram for each wind speed sequence i. In considering all the measurements sequences together, then $X_{*l}$ gives a vector of samples for bin l. To simplify the notation, when we write $X_l$ we imply $X_{*l}$.

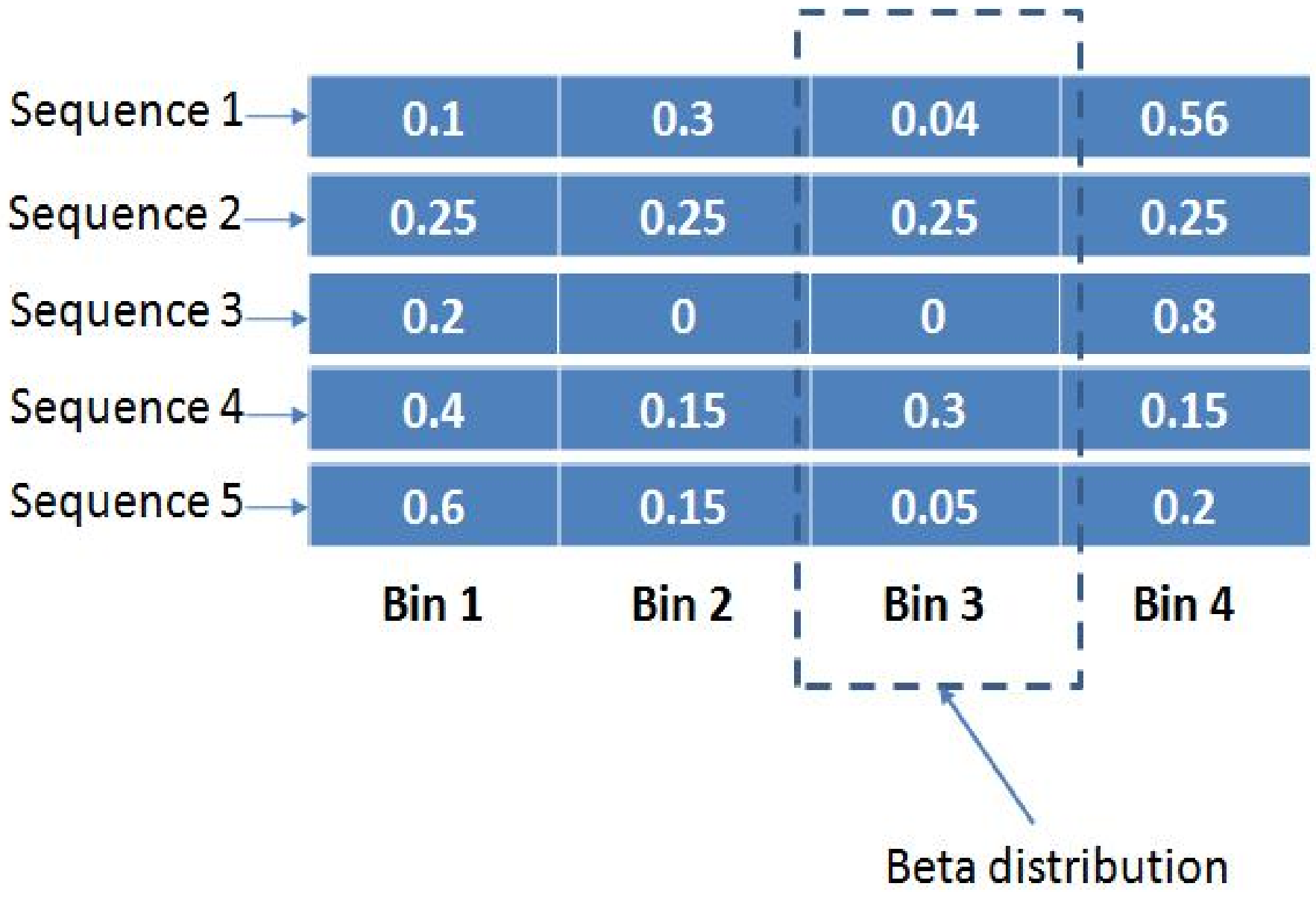

***Figure 1: An example with 5 sequences and 4 bins.***

The vector $X_l$ gives a set of samples on the proportion of time that an arbitrary sequence has speed values in the range defined by the l-th bin. An example of this vector is indicated in Figure 1 via the encircled set of values. We can thus define the vector $\chi = (X_1, ..., X_L)$ (e.g., a vector of vectors) to represent our entire data collection.

In Figure 2 we plot three signals of wind speed sequence and the corresponding histograms they generate. These three measurements sequence generate three different histograms, in other terms three different wind regimes.

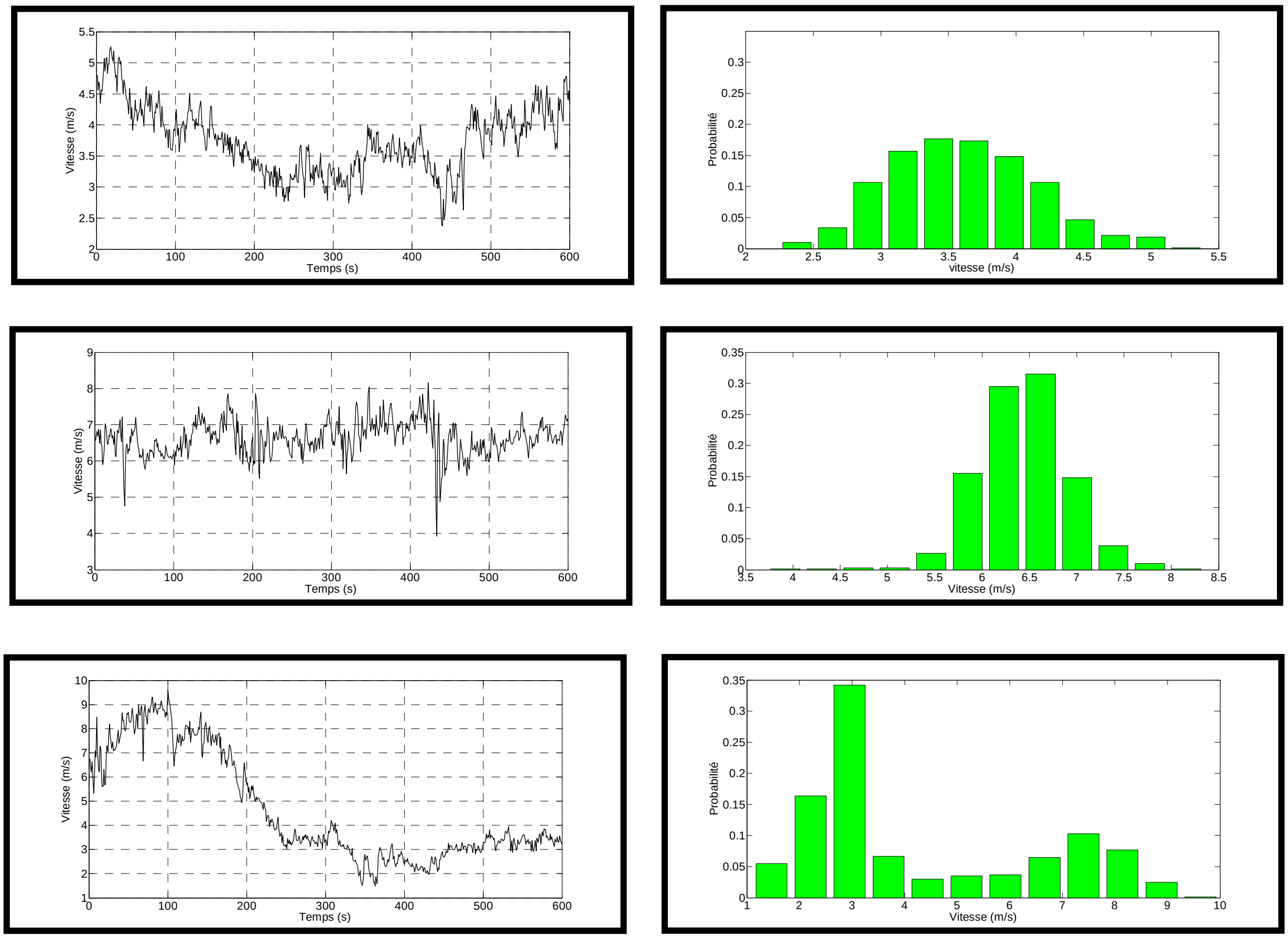

**Figure 2: Three examples of wind regimes and the corresponding histograms.**

### 3.2 Motivation

In order to construct forecasting tools, a classification and an identification of wind speed sequences is necessary. In [2], the analysis of the statistical of the first second moments (mean, standard deviation) evaluated from the wind speed samples of each time sequence, is presented. The whole set of measurement has been divided in consecutive sequences of duration T = 10 minutes. For each of these sequences, the mean value $\overline{U}$ and the standard deviation σ of wind speed are computed. In figure 3, the standard deviation σ was plotted versus the mean wind speed $\overline{U}$. Each point of the map ($\overline{U}$, σ), represents a sequence of 10 minutes wind speed samples.

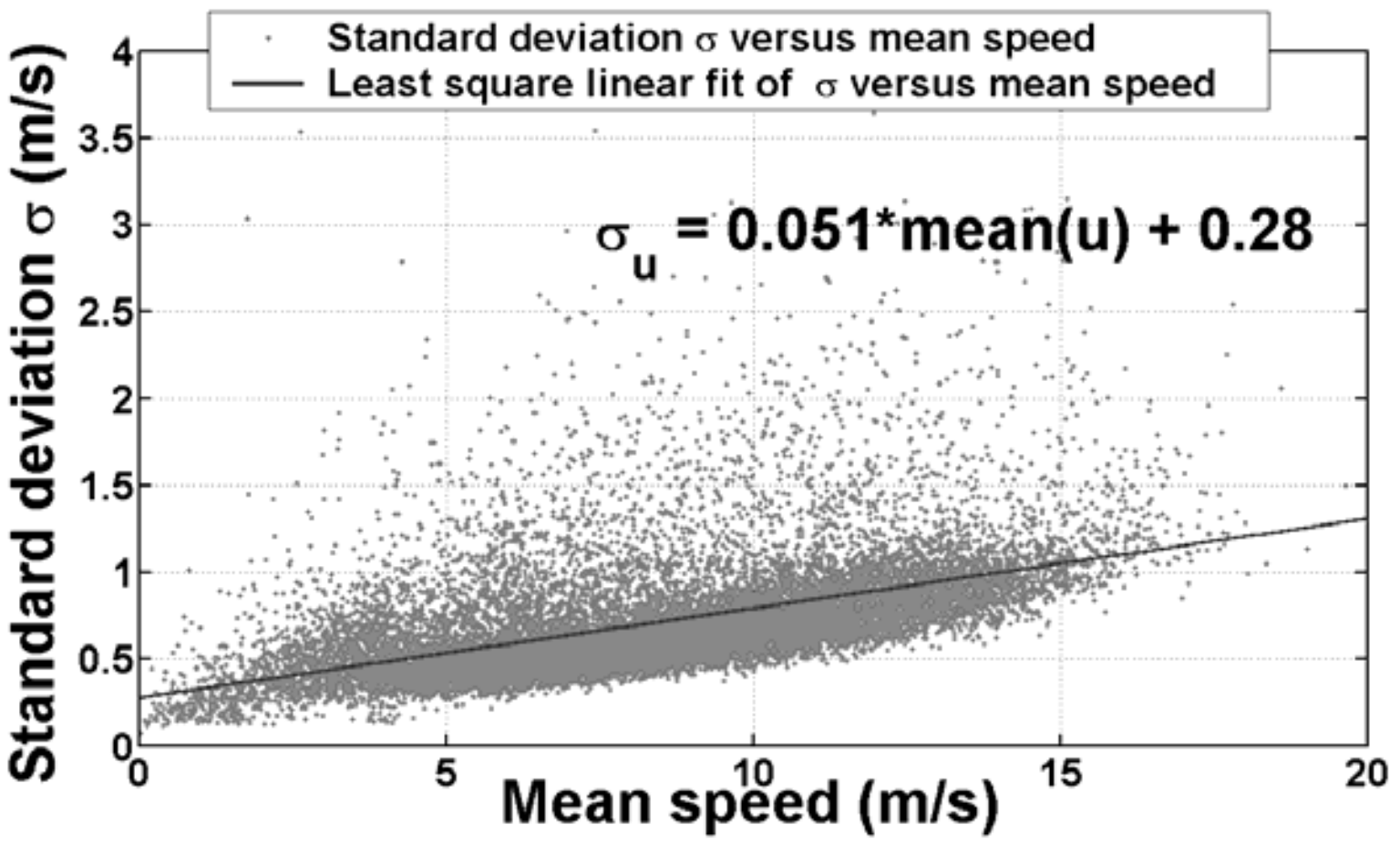

**Figure 3: Standard deviation as a function of average wind speed for averaging time *N*=600 s.**

In [2], each of the experimental distribution has been compared to the normal law distribution on the basis of the Kolmogorov – Smirnov parametric test for goodness of fit. The results have shown that in the vicinity of the least square linear fit, at least 80% of the experimental wind sequences can be considered as having a Gaussian PDF. Concerning, the time sequences for which a Gaussian distribution cannot be used, the shapes of the experimental PDF are more complex. Some asymmetrical mono-modal PDF are observed in regions well above the least square linear fit and bimodal PDF are observed for σ $\geq 2$ m/s: the Kolmogorov – Smirnov test doesn't show the existence of these PDF.

The classification method developed in this paper, allows to highlights the different experimental wind speed PDF observed.

## IV. Methodology

In our approach, any vector $\chi = (X_1,..., X_L)$ that represents the empirical histogram of wind speed sequence, *L* being a number of bins, will be considered as an outcome of a random variable whose distribution is a finite mixture of Dirichlet distributions. To explain this, we first present some background material on Dirichlet distributions and then we present the algorithm that estimates the finite mixture.

**4.1 Theoretical framework**

Let $V = \{1,..., L\}$, and let $P(V)$ denote the set of all probability measures defined over the finite set $V$, so that $P(V)$ can be identified to the set $S_L = \{x_1,..., x_L\}$ with $x_i \geq 0$ and $\sum_{i=1}^{L} x_i = 1$. Note that any observed histogram belongs to $S_L$.

This last random distribution (RD) is very useful in the context we are studying as each histogram coming from a flow is a discrete probability distribution defined over a finite set of bins. Remember that $X_l$ is a random variable that denotes the likelihood of a wind speed value in the range defined by bin *l*. Therefore the last example of RD describes the set of histograms we have to deal with in flow classification. The source generating random distributions is governed by a multidimensional probability distribution that jointly defines the probability of an histogram $\chi = (X_1,..., X_L)$. Clearly bin sizes $(X_i)$ are dependent of each other as they are jointly constrained by the condition that $\sum_{l=1}^{L} X_K = 1$.

Let B the Dirichlet distribution density, with parameter vector $\alpha = (\alpha_1,........., \alpha_L)$ is given by

$$f(x_1, x_2,...., x_L / \alpha_1, \alpha_2,...., \alpha_L) = \frac{\Gamma(\alpha_1 + ... + \alpha_L)}{\Gamma(\alpha_1)... \Gamma(\alpha_L)} \prod_{l=1}^{L} x_l^{\alpha_l - 1} \left(1 - \sum_{l=1}^{L-1} x_l\right)^{\alpha_l - 1}$$

where $\alpha_1 > 0,....,\alpha_L > 0,....,x_l > 0$, and $\sum x_l = 1$.

This defines the joint density probability function of $(X_1 = x_1, X_2 = x_2,..., X_L = x_L)$. We denote this Dirichlet distribution by $D(\alpha_1,...,\alpha_L)$.

Remarks:

The popularity of the Dirichlet distribution is due to several convenient properties listed here:

1. The Dirichlet distribution can be simulated easily by the following normalization construction. Suppose $Z_1,....,Z_L$ are $L$ random variables following gamma distributions $\gamma(\alpha_1,1),....,\gamma(\alpha_L,1)$[2] respectively, where

$$\gamma(a,b)(x) = \frac{1}{\Gamma(a)} b^a e^{-bx} x^{a-1} I_{(O,+\infty)}(x)dx \text{ with } x > 0$$

If we normalize each random variable $Z_l$ by the sum $Z_l = Z_1 + ... + Z_l$, then $Z_l/Z$ has a beta distribution, and the multivariate random vector $\left(\frac{Z_1}{Z},...,\frac{Z_L}{Z}\right)$ will follow a Dirichlet distribution $D(\alpha_1,...,\alpha_L)$. Because we try to analyze the histogram of a wind speed sequence rather than its values, we are dealing with observations that are themselves probability distributions. In other words, each yields a histogram that can be seen as a realization coming from a stochastic source generating random histograms. To make things more clear we can note that a random variable represents a source that generates a single value, a random vector represents a source that generate vectorial observations, and in our case as the source generates histograms we have to deal with random distribution. A formal definition of random distributions is given as follows. An example of a random distribution, for discrete random variables, is the following.

2. The Dirichlet distribution has the following nice property that is particularly useful. If $\chi = (X_1,..., X_L)$. has a Dirichlet distribution, $D(\alpha_1,...,\alpha_L)$ then the marginal distribution of each component $X_l$ follows a beta distribution: $X_l B \approx (\alpha_l, A-\alpha_l)$ where A, defined as $A = \sum_{l=1}^{L} \alpha_l$ _l, is called the mass-value. The mean is given by $\mathrm{E}\{X_l\} = \frac{\alpha_l}{\sum_{l=1}^{L} \alpha_l}$ and the variance is given by $\mathrm{Var}(X_l) = \frac{\alpha_l (A-\alpha_l)}{A(A+1)}$. In other words the variance of all components $X_l$ is governed by the mass-value $A$.

3. We should notice that the problem of estimating the real distribution of a flow based on an observed empirical distribution over k bins can be formalized as estimating the parameters $\Theta = (\theta_1,......,\theta_L)$ of a multinomial distribution based on an observed empirical histogram $\chi = (X_1,..., X_L)$. Now if a random variable *Z* follows a multinomial distribution $\mathrm{M}(\theta_1,....,\theta_L)$ with unknown parameters $\Theta = (\theta_1,......,\theta_L)$ and if the prior distribution on the unknown parameter $_\pi(\Theta)$ is a Dirichlet distribution $D(\alpha_1,...,\alpha_L)$ , the posterior probability $\mathrm{Prob}\{\Theta / X = (x_1,....,x_L)\}$ will also follow a Dirichlet distribution given by $D(\alpha_1 + x_1,...,\alpha_L + x_L)$, i.e. the prior distribution has the same form as the posterior distribution. In other words the Dirichlet distribution is the conjugate prior for the multinomial distributions. This property reduces the updating of the prior based on the observed value, to a simple update of the parameters in the prior density. It is therefore natural to use a Dirichlet distribution in the context of inference of a finite distribution.

All these three properties make the Dirichlet distribution very attractive for modelling random distributions. Moreover Dirichlet distributions, and more specially the mixtures of Dirichlet distributions (to be defined later in the paper), have demonstrated, in practice, a good ability to model a very large spectrum of different distributions observed in the real world [5], [17], [6].

**4.2 Mixture of Dirichlet distribution**

Mixed Dirichlet Dristribution (MDD) is often used as a flexible and practical way for modelling prior distributions in nonparametric Bayesian estimation. The rationale for using Dirichlet mixtures is well explained in [17]. Examples of applications include empirical Bayes problems [5], nonparametric regression [13] and density estimation [6]. In this paper we want to classify observed wind speed sequences based on the similarity of their distribution. We assume that the observed empirical histograms are coming from a source governed by a MDD. Rather than finding a single distribution to represent all flows, it makes intuitive sense to think of each class of time series as having its own distribution. The entire ensemble of n time series $\chi = \{X_{i*}, i = 1, i = 1, \ldots.., n\}$, that contains the empirical histogram for each flow, is modelled as a mixture of multiple Dirichlet processes defined as $\sum_{k=1}^{K} p_k D(\alpha_1^k, \ldots.., \alpha_L^k)$. where each component $D(\alpha_1^k, \ldots, \alpha_L^k)$ represents a time series class, and each $p_k$ represents the weight assigned to the class. This mixture defines the so called a priori probability of the class. Now each observed histogram is assumed to come from one of these components. The classification problem consists of determining from which source component each histogram could have originated. To solve this problem, we need to find out the a posteriori probability, i.e. the probability that a wind speed sequence belongs to a class given the histogram of the flow. MDDs inherits the nice properties of Dirichlet processes we described in the previous section. Any particular probability density can be approximated over a bin set $B$ by a MDD with suitable parameters. Moreover the mass-value of each component controls the extent to which the model is allowed to diverge from its specified mean behaviour. So MDD doesn't contain as much a priori as a normal or Poisson distribution.

Let $K$ denote the number of classes into which we want to classify our wind speed sequences. We model our observed histograms by assuming that the distribution of bins $P_r(X_1,...,X_L)$ can be described by a finite mixture of K Dirichlet distributions:

$$P_r(X_1,...,X_L)=\sum_{k=1}^{K} p_k D(\alpha_1^k,....,\alpha_L^k)$$

where the coefficients $p_1,...,$ $p_k$ denote the weight, or contribution, of each Dirichlet density. This gives the prior distribution, that is the probability that one observes $(x_1,....,x_L)$ given that the parameters are fixed at $p_1,....,p_K$ and $\alpha_1^k,....,\alpha_L^k$ for $k=1,....,K$. However in practice these parameters are unknown and in order to finalize our model, we need to estimate them. Based on this a priori probability e need also to obtain the a posteriori or the class membership probability, i.e. the probability that a flow belongs to a class given the histogram of the flow. In the following section we present the estimation procedure for estimating these parameters based on our data.

### 3.3 Estimation procedure

Several methods have been proposed to estimate the mixing weights $p_k$ and the parameters of the components $P_k$; here we use one of the most efficient methods called SAEM, a Simulated Annealing Expectation Maximization algorithm [14]. SAEM is a stochastic approximation of the popular Expectation Maximization (EM) algorithm [11] that is less sensitive to local minima problems. The EM algorithm is a general method of finding the maximum likelihood estimates of the parameters of an underlying distribution from a given data set when the data is incomplete or has some unknown parameters. The EM method is based on iteration between Estimation and a Maximization step. The usage of the EM algorithm in the case of mixture models is well described in [1].

SAEM, as first described by Celeux and Dielbot in [3], modifies the EM methods to get rid of common problems encountered such as slow convergence or local maxima. Instead of using a prior distribution for the unknown parameter it involves a stochastic step that simulates the unknown data in order to obtain complete data and to uncover hidden variables.

Our algorithm takes as inputs the histograms, the number of desired classes $K$, and a sequence of values $\gamma_q$. These values $\gamma_q$ are used to control the trade-off between the influence of the stochastic step and the EM steps. Let $\{q\}$ be a sequence of positive real numbers decreasing to zero at a sufficiently slow rate, with $\gamma_0 = 1$. Each time the algorithm iterates, repeating the E and M steps, the impact of the stochastic EM component is successively reduced (by multiplying with smaller and smaller $\gamma_q$). When $\gamma_q$ approaches zero, our algorithm reduces to a pure EM algorithm.

Our algorithm outputs three things: the weights $p_k$, of each Dirichlet process; the Dirichlet parameters $\alpha = (\alpha_1, ..., \alpha_L)$ and the class membership probabilities $t_{ik}^q$, where $t_{ik}^q$ denotes the probability that wind speed sequence histogram $i$ belongs to class $k$ at the $q^{th}$ iteration of the algorithm. This algorithm asymptotically estimates the parameter of the mixture model since $p_{qk}$, $t_{ik}^q$ and the density parameters converge as $q \to \infty$ [3]. A general formulation of the SAEM for the large class of mixtures of density functions belonging to the exponential family has the form:

$$d(x,a) = d^{-1}(a)e(x)\exp < a^T.b(x) >$$

where the parameter $a$ a is a vector with transpose $a^T$ , $d(a)$ is a normalizing factor, $e$ and $b$ are fixed but arbitrary functions and < . > is the standard inner product. In adapting this to our problem, the case of Dirichlet mixtures, we need to set the parameters as follows, $a = (\alpha_1,...,\alpha_L)$, $b(x) = (\log(x_1),...,\log(x_L))$, $d(a) = \frac{\Gamma(\alpha_1)...\Gamma(\alpha_L)}{\Gamma(\alpha_1 + ... + \alpha_L)}$ and $e(x) = x_1^{-1}...x_L^{-1}$1. The inputs are the n vectors $X_{i*}$ $i = 1,...,n$ where each observation $X_{i*}$ is a normalized histogram. The number of components in the mixture is a given integer $K$ assumed to be known. Our algorithm is given; this algorithm contains three main steps:

• A simulation step that introduces some noise into the process by making a random class assignment. This noise helps pushing the algorithm out from local minima. However since the parameter $\gamma_q$ is decreasing, the noise decreases as well, and the algorithm will converge to a stable estimate. A threshold $c(n)$ is used where $0 < c(n) < 1$ and $\lim_{n\to\infty} c(n) = 0$. This threshold determines whether or not one needs to return to the initialization step and essentially start over.

• A maximization step that updates the parameter values $a_k^{q+1}$, as well as the mixing weights $p_{(q+1)k}$, such that the likelihood is maximized. (Recall that the $a_k^{q+1}$ variables in the algorithm correspond to the $\alpha$ variables in our model as stated above.)

• An estimation step in which we update the membership probabilities $t_{ik}^q$, i.e., the probability that wind speed sequence $i$ belongs to class $k$ (at the $q^{th}$ iteration through the algorithm). Recall that this is our posterior distribution (in Bayesian terms).

*Initialization step:*

*Assign randomly each wind sequence i to a class.*

*Simulation step:*

*Generate randomly* $t_{ik}^{(0)}(i=1,...,n)$ *representing the initial a posteriori probability that a wind speed sequence* $i$ *is in class* $k$ *where* $1 \le k \le K$ *.*

*For* $q=0$ *to* $Q$ *do*

*Stochastic step:*

*Generate random multinomial numbers* $e_{qi} = \left(e_{qi}^{k}\right)$ *following the probability distribution* $\left\{t_{ik}^{q}\right\}$ *where all the* $e_{qi}^{k}$ *are 0 except one of them equal to 1. We then get a partition* $C = (C_k)_{k=1,...,K}$ of the set of histograms

*If* $\frac{\sum_{i=1,...,N} e_{qi}^{k}}{N} < c(n)$ *for some* $k$ *then*

*Return back to initialisation step.*

*End*

*Maximisation step:*

*Estimate the mixing weights* $p_k^{(q+1)} = \frac{1}{n}\left[\left(1-\gamma_q\right)\sum_{i=1,...,n} t_{ik}^{q} + \gamma_q \sum_{i=1,...,n} e_{qi}^{k}\right]$

*and the parameter value* $a_k^{q+1} = \left(1-\gamma_q\right)\frac{\sum_{i=1,...,n} t_{ik}^{q} b(f_i)}{\sum_{i=1,...,n} t_{ik}^{q}} + \gamma_q \frac{\sum_{i=1,...,n} e_{qi}^{k} b(f_i)}{\sum_{i=1,...,n} e_{qi}^{k}}$

*Estimation step:*

*Update the a posteriori probability of a histogram* $i$ *belonging to class*

$$k\left(t_{ik}^{q+1}\right) = \frac{p_{(q+1)k} D\left(\alpha_{1,q+1}^{k},...,\alpha_{l,q+1}^{k}\right)(f_i)}{\sum_{r=1...K} p_{(q+1)} D\left(\alpha_{1l,q+1}^{k},...,\alpha_{l,q+1}^{k}\right) h_{(q+1)r}(f_i)}$$

*End*

**Algorithm 1: SAEM algorithm.**

# 5. Results

The data used for this classification study was described in section 2. Recall that we have approximately 1,000,118 wind speed sequences of duration 10 minutes with sample rate of 1 second. From these sequences, we have constructed the histograms of wind speed sequences. These distributions contain 12 bins: the choice of the number of bins and the location of the bin centers $(B)$ is important. One the one hand the larger the number of bins the more accurately our empirical histogram will represent the real distribution. One the other hand, if there are too many bins, some bins might remain empty and the estimation algorithm might fail (because an empty bin gives a likelihood of zero). In order to follow these points, our histograms have 15 bins.

The proposed algorithm is applied on the ensemble of the wind speed sequences in order to find $K$ classes that represent the ensemble of all of these histograms. This number $K$ thus also defines the number of Dirichlet processes in the mixture model.

In first time, we present the results of the classification with two classes, after with three classes.

## 5.1 Classification with two classes

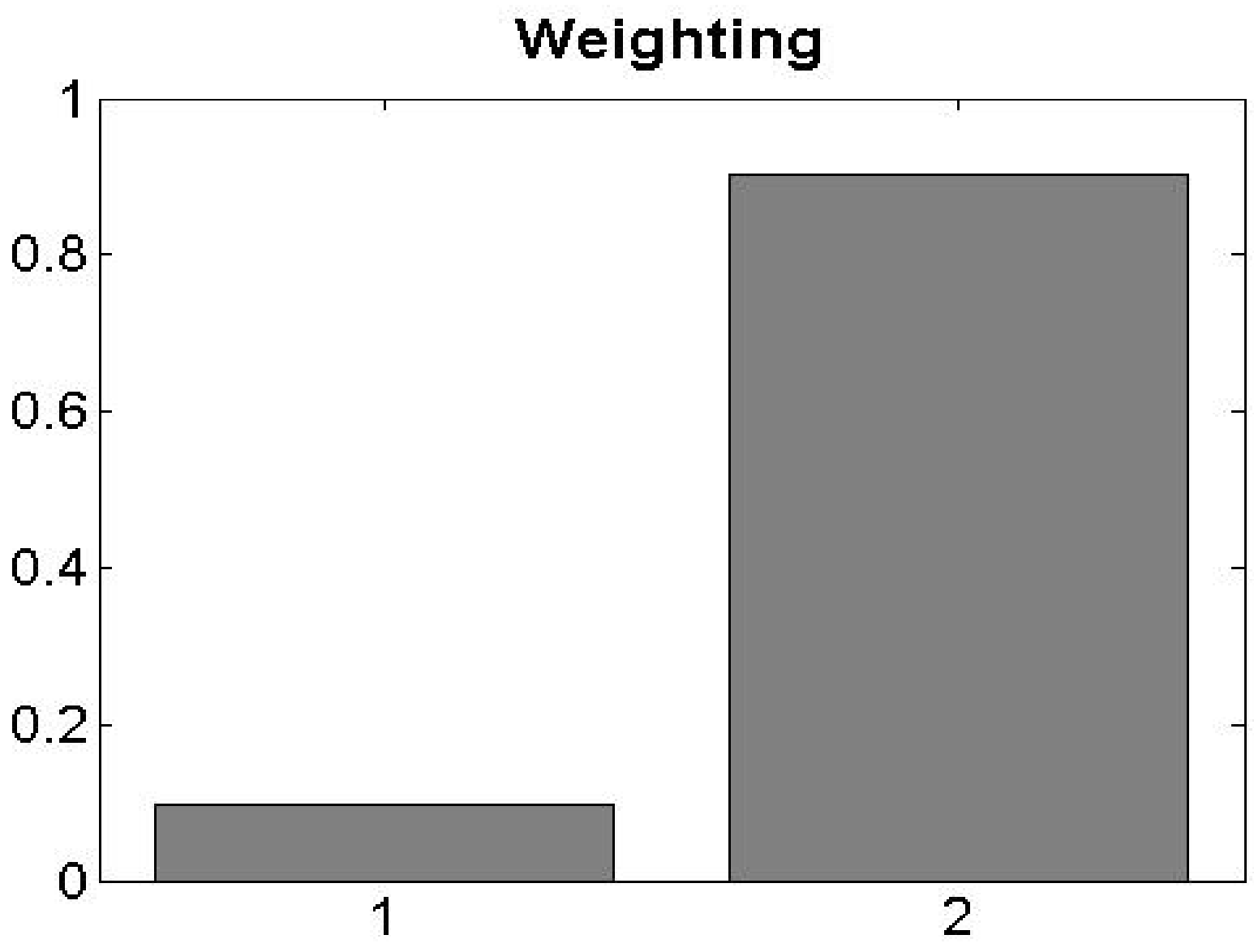

**Figure 4: Weighting of each class.**

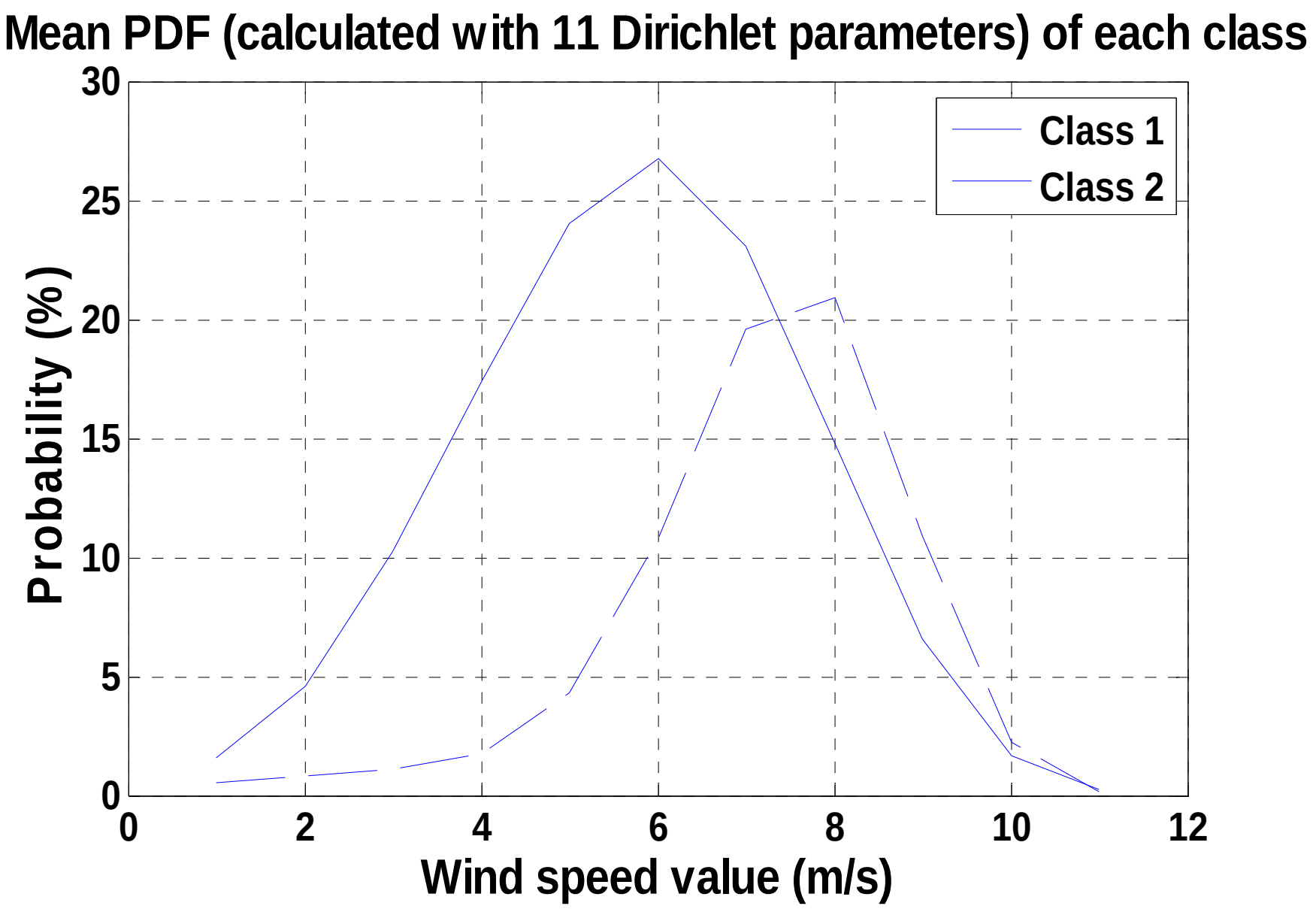

**Figure 5: Mean class PDF of each class.**

Here we classify all the wind speed sequences into two classes. Applying the SAEM algorithm, we found that 100,011 wind speed sequences $(10\%)$ belong to class 1 and that 900,107 wind speed sequences $(90\%)$ are classified as class 2. Figure 4 presents the distribution of the mean for each class calculated with 11 Dirichlet parameters. The mean behaviour of class 2 follows a symmetrical mono-modal PDF for the wind speed distribution.

In this class, the standard deviation mean is equal to $0{,}67m/s$. Concerning the class 1, the mean behaviour follows an asymmetrical mono-modal PDF for the wind speed distribution. This PDF characterize wind speed sequences that the standard deviation mean is equal to $0{,}89m/s$. A point of view meteorological, these times series are strong wind regimes. Figure 6 illustrates two examples of wind speed sequence belong to these classes.

This empirical classification with the number of desired of classes $K = 2$, highlights the principals classes of the wind speed sequences, i.e., the sequences corresponding to some of the points included in the region of the map $(\overline{U},\sigma)$ positioned in the vicinity of the least square linear plot fit in figure 2.

In order to verify the existence of another class, we classify wind speed sequences using both three classes.

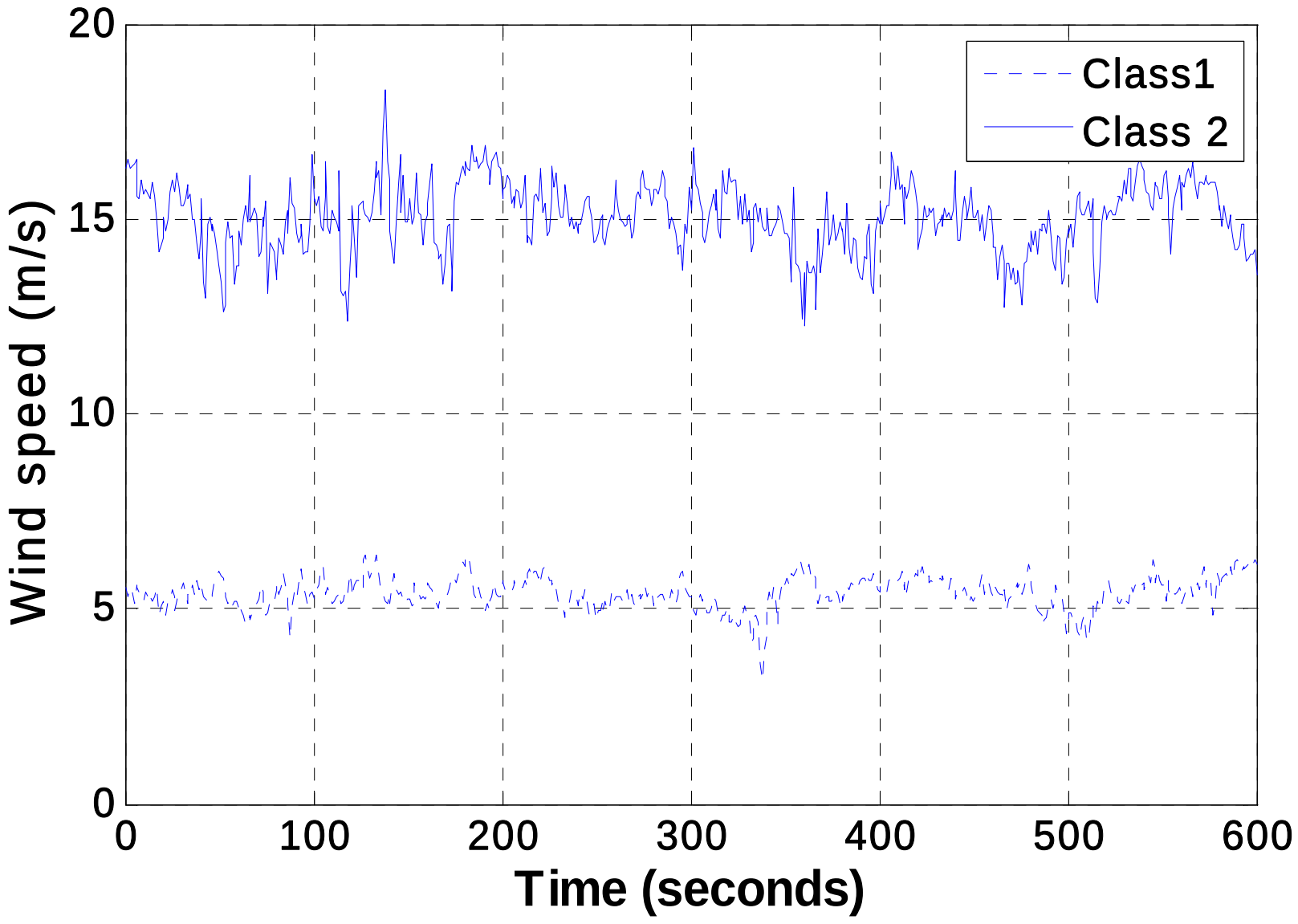

**Figure 6: Two examples of time series for each class.**

### 5.2 Classification with three classes

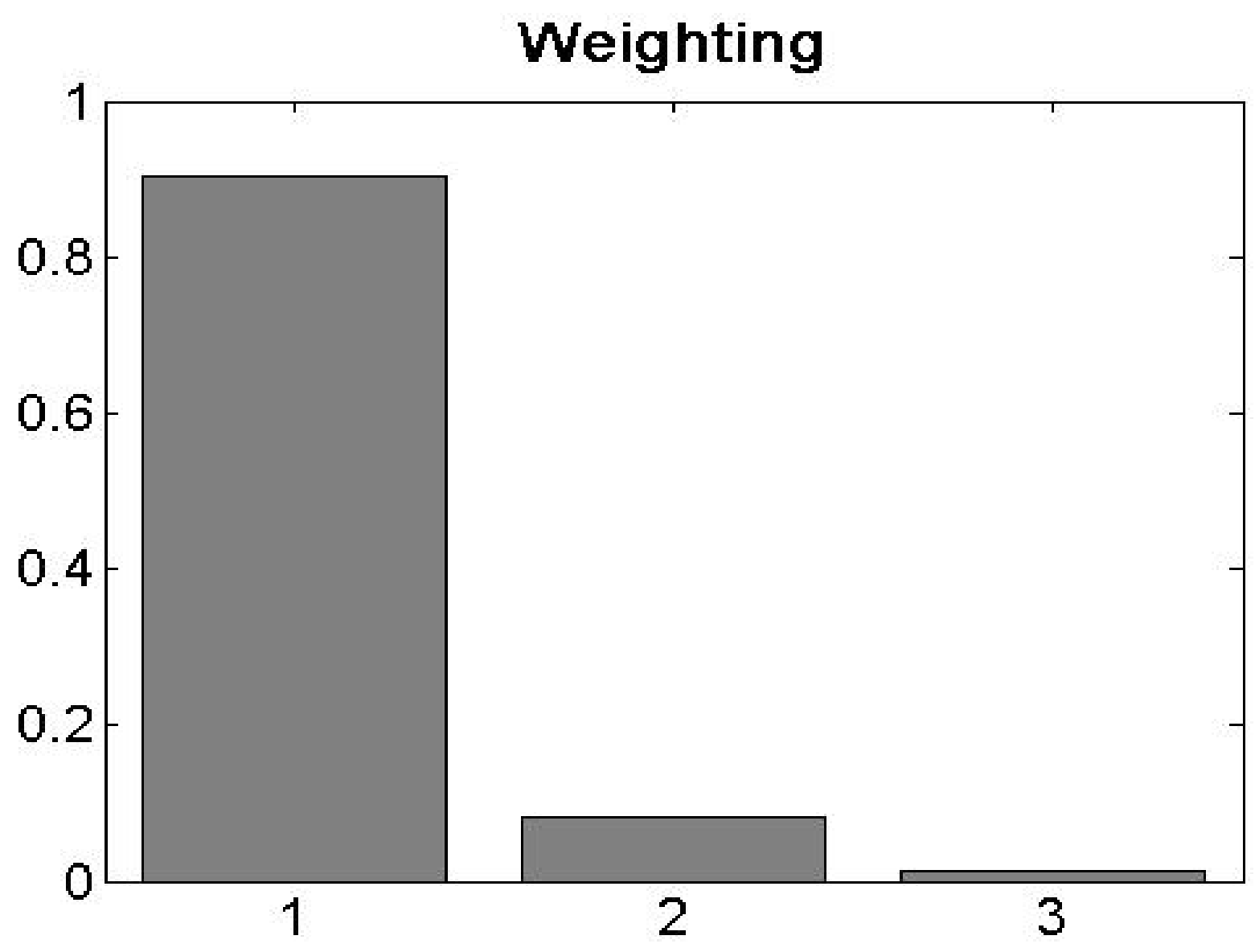

**Figure 7: Weighting of each class.**

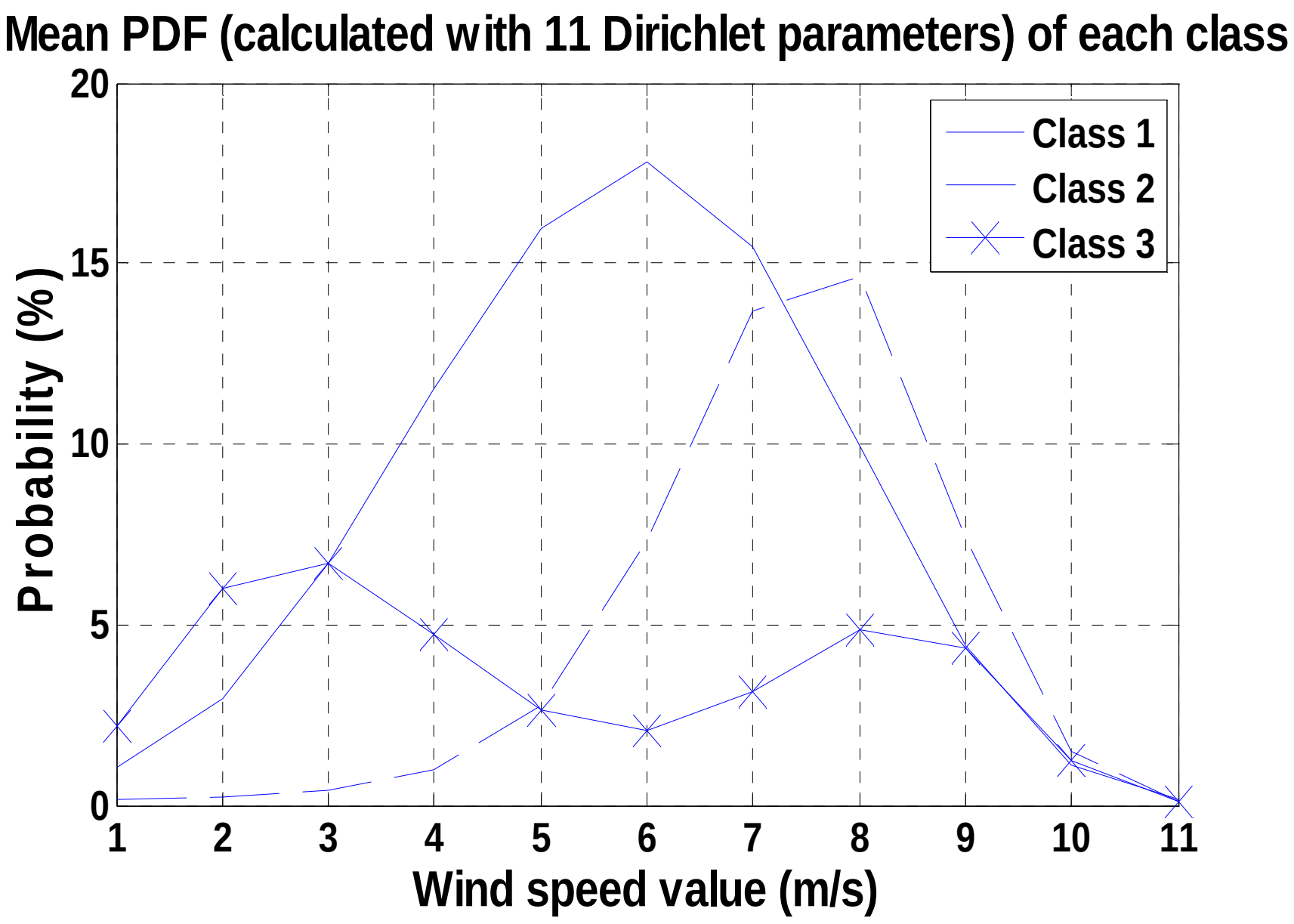

**Figure 8: Mean PDF of each class.**

In figure 6, we illustrate the CDF and the PDF of each class. The results of classification give the two first previous classes with a new class. Indeed, 900,101 wind speed sequences $(90\%)$ are classified in class 90,010 wind speed sequences $(9\%)$ belong to class 2 and 10,001 wind speed sequences $(1\%)$ belong to class 3. The two first classes are the classes found previously. Concerning, the last class, the mean behaviour follows a bimodal PDF for the wind speed distribution. This bimodal PDF characterize wind speed sequences that the standard deviation calculated over 10 minutes period, is superior to $1.5m/s$; indeed the standard deviation mean in this class, is equal to $1{,}98m/s$. Moreover, these measurement sequences whose signatures in the map $(\overline{U},\sigma)$ (figure 2) are located far from the linear fit. We notice for these marginal measurement sequences, the value of turbulent intensity is greater than $20\%$ [2], which means a strong turbulent agitation. We can suppose that these measurement sequences can correspond to gust of wind or heavy shower. These marginal sequences are very drastically for wind energy production.

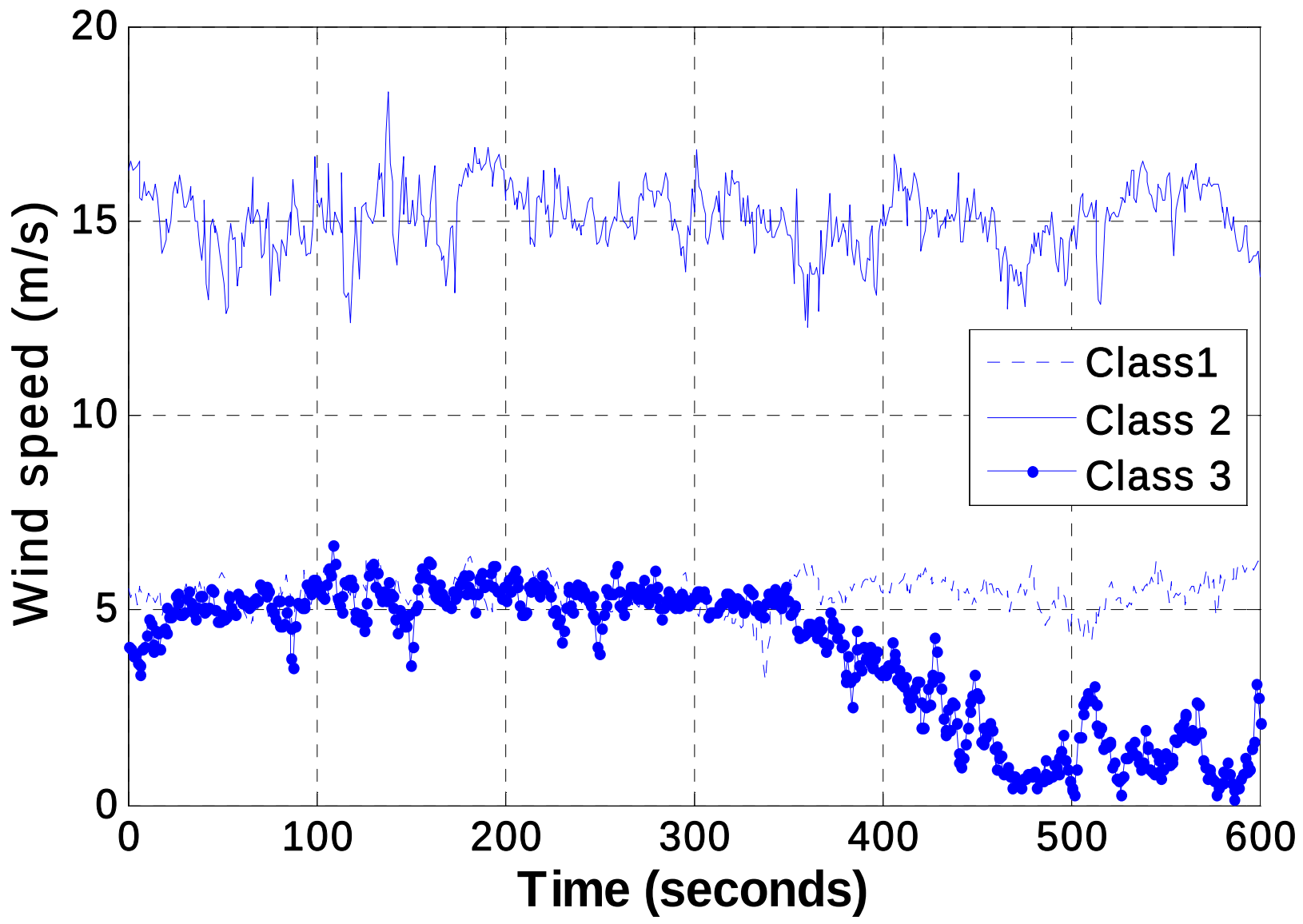

**Figure 7: Example of three time series for each class.**

### 5.3 Parametric model for the wind speed distribution

*a) A Gram-Charlier PDF*

Besides, we propose a mathematical function called Gram-Charlier that can be used to model the wind speed distribution in class 1 and class 2. The general form of the Gram-Charlier density is given by [9], [10]:

$$g(u) = p_n(u)\phi(u)$$

Where $u$ represents the value of the wind speed, $\phi(u)$ a Gaussian distribution and $p_n(u)$ a polynomial developed in the Hermite polynomial basis:

$$p_n(u) = \sum_{i=0}^{n} c_i He_i(u)$$

Where $He_i(u)$ are the Hermite polynomials. A statistical notation in the literature is

$$p_4(u) = 1 + \frac{s}{6} He_3(u) + \frac{k}{24} He_4(u)$$

where s is the skewness coefficient and k the kurtosis coefficient. This case corresponds to the Gram-Charlier type-A and the Edgeworth expansions [15].

This function is used to model the wind speed distribution for the two classes:

1) Wind speed distribution in the first class is well modeled by a normal density or a Gram-Charlier density with $p_n(u) = 1$ and $\phi(u)$ a Gaussian distribution. Figure 10 illustrates a wind speed distribution approximated by a normal density, which has same mean value $\overline{U}$ and same standard deviation $\sigma$ as the wind speed sequence.

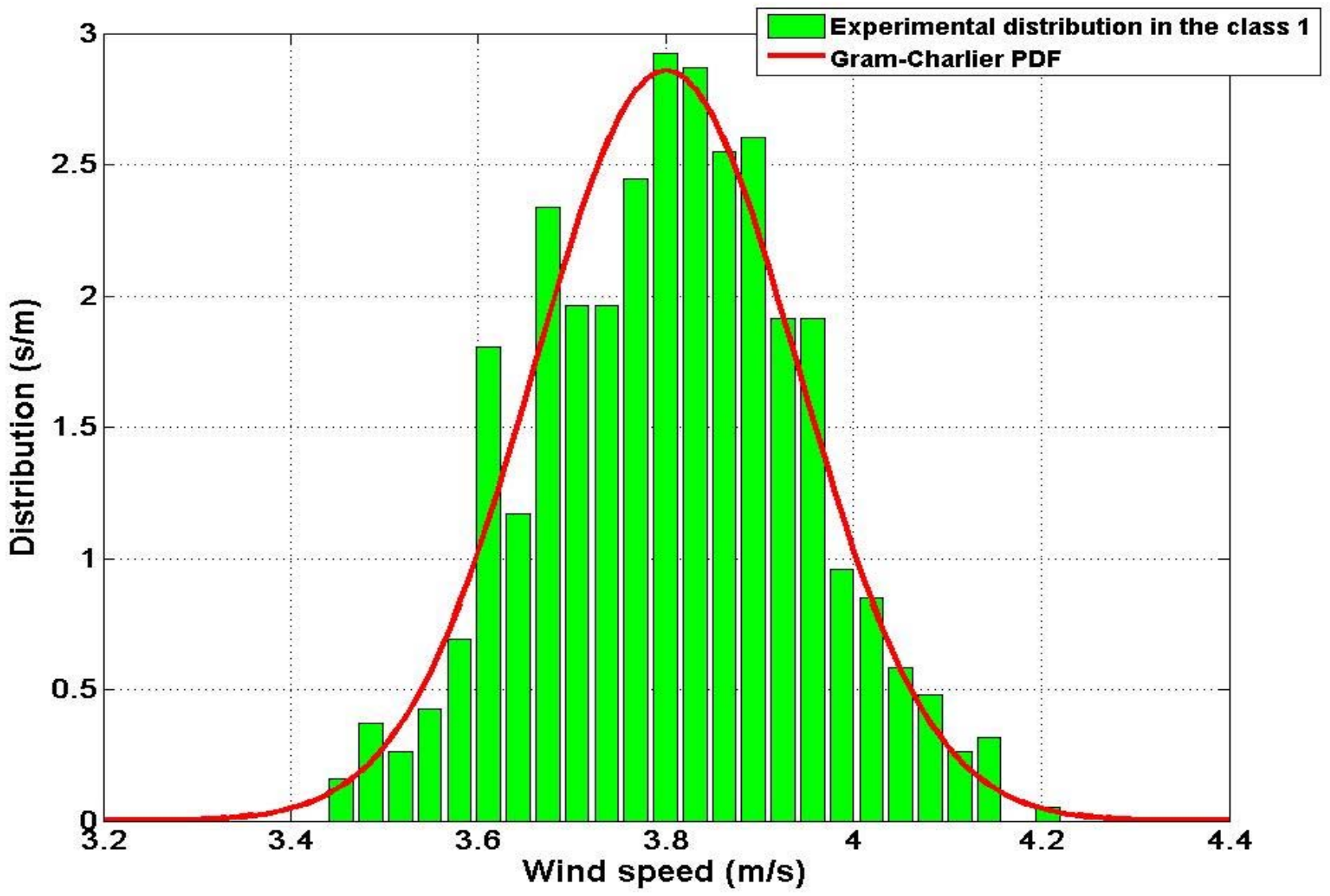

**Figure 10: Wind speed distribution in class 1**

2) Wind speed distribution in the second class, is approximated by a Gram-Charlier function, with $\phi(u)$ a Gaussian distribution and $p_4(u)=1+\frac{s}{6}He_3(u)+\frac{k}{24}He_4(u)$. In this case, the mean value $\overline{U}$ , the standard deviation $\sigma$, the skewness coefficient s and the kurtosis coefficient k, represent the four first moments of a wind speed sequence in the second class. Figure 11 shows a wind speed distribution modeled by a Gram-Charlier type-A.

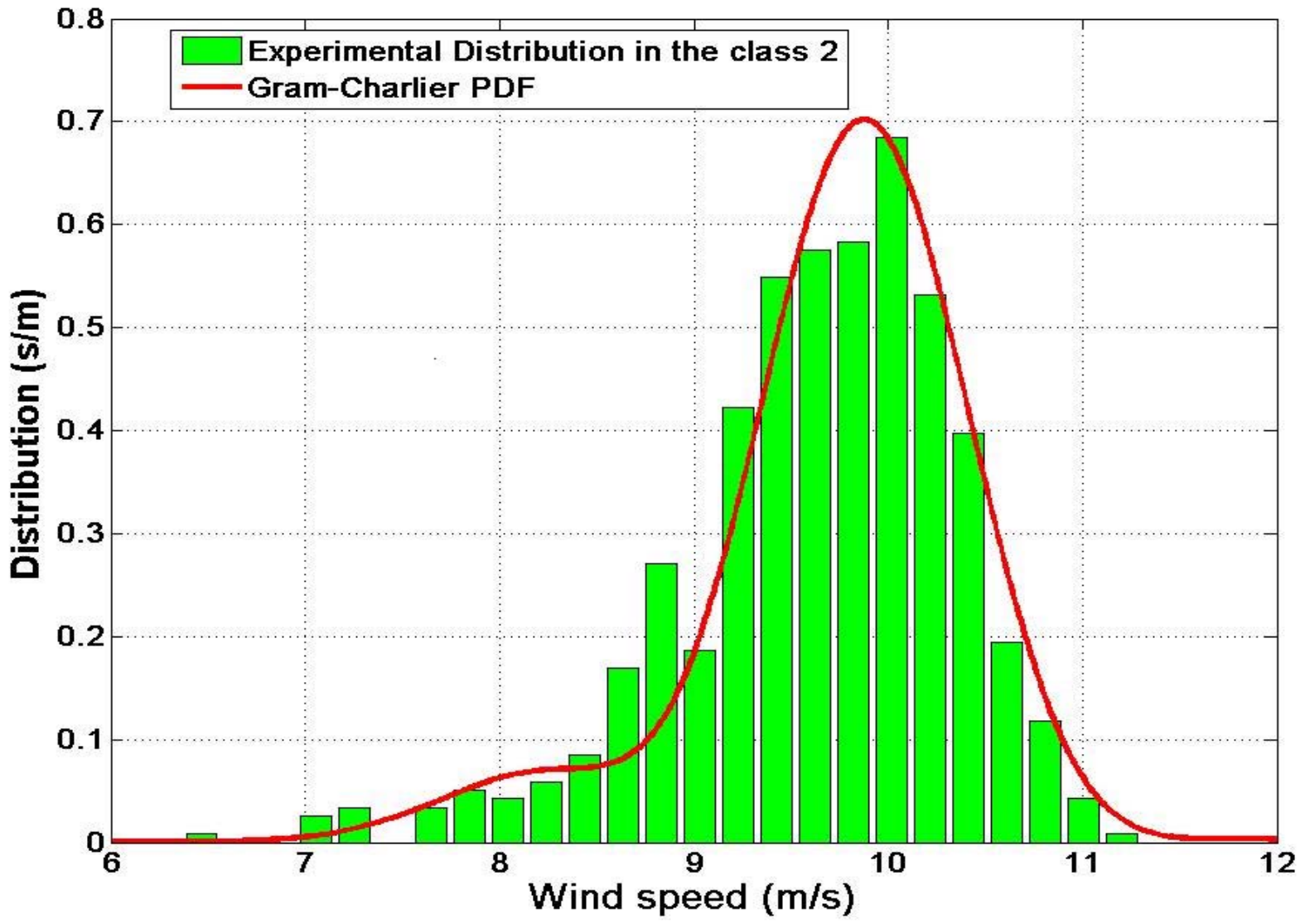

**Figure 11:. Wind speed distribution in class 2.**

*b) A bi-Weibull PDF*

In our analysis, we considered a bimodal PDF to fit the wind speed PDF in class 3. This function is based on a Weibull&Weibull PDF [20]:

$$F_{WW}(u) = p[F_W(u)]_{left} + (1-p)[F_W(u)]_{right}$$

Or in an explicit manner

$$F_{WW}(u) = p\int_0^\infty \frac{k_1}{c_1}\left(\frac{u}{c_1}\right)^{k_1-1} \exp\left[-\left(\frac{u}{c_1}\right)^{k_1}\right]du + (1-p)\int_0^\infty \frac{k_2}{c_2}\left(\frac{u}{c_2}\right)^{k_2-1} \exp\left[-\left(\frac{u}{c_2}\right)^{k_2}\right]du = 1$$

where $F_{WW}(u)$ is the bimodal Weibull&Weibull PDF, $u$ is the wind speed, $c_1$ and $c_2$ are the scale parameters established by the left and right Weibull distribution, respectively; $k_1$ and $k_2$ are the shape parameters established by the left and right Weibull distribution, respectively;

and $p$ is the weight component of the left Weibull distribution (0<$p$<1). The weight component $p$ can be obtained by using the following formulas,

$$\overline{U} = p\overline{U}_1 + (1-p)\overline{U}_2$$

And

$$\sigma^2 = p\left(\sigma_1^2 - (p-1)(\overline{U}_1 - \overline{U}_2)^2\right) - (p-1)\sigma_2^2$$

Where $\overline{U}$ is the average wind speed sequence and $\sigma$ is the standard deviation the wind speed sequence; $\overline{U}_1$ and $\overline{U}_2$ are the average wind speed sequence of the left and right Weibull distribution, respectively; $\sigma_1^2$ and $\sigma_2^2$ are the variance of the left and right Weibull distribution. The parameters $c_1$, $c_2$, $k_1$, and $k_2$ can be obtained by solving the equations,

$$\overline{U}_i = c_i \Gamma\left(1 + \frac{1}{k_i}\right)$$

And

$$\sigma_i^2 = c_i^2\left[\Gamma\left(1 + \frac{2}{k_i}\right) - \Gamma^2\left(1 + \frac{1}{k_i}\right)\right]$$

Where $I$ =1 for the left Weibull distribution and $I$ =2 for the right Weibull distribution.

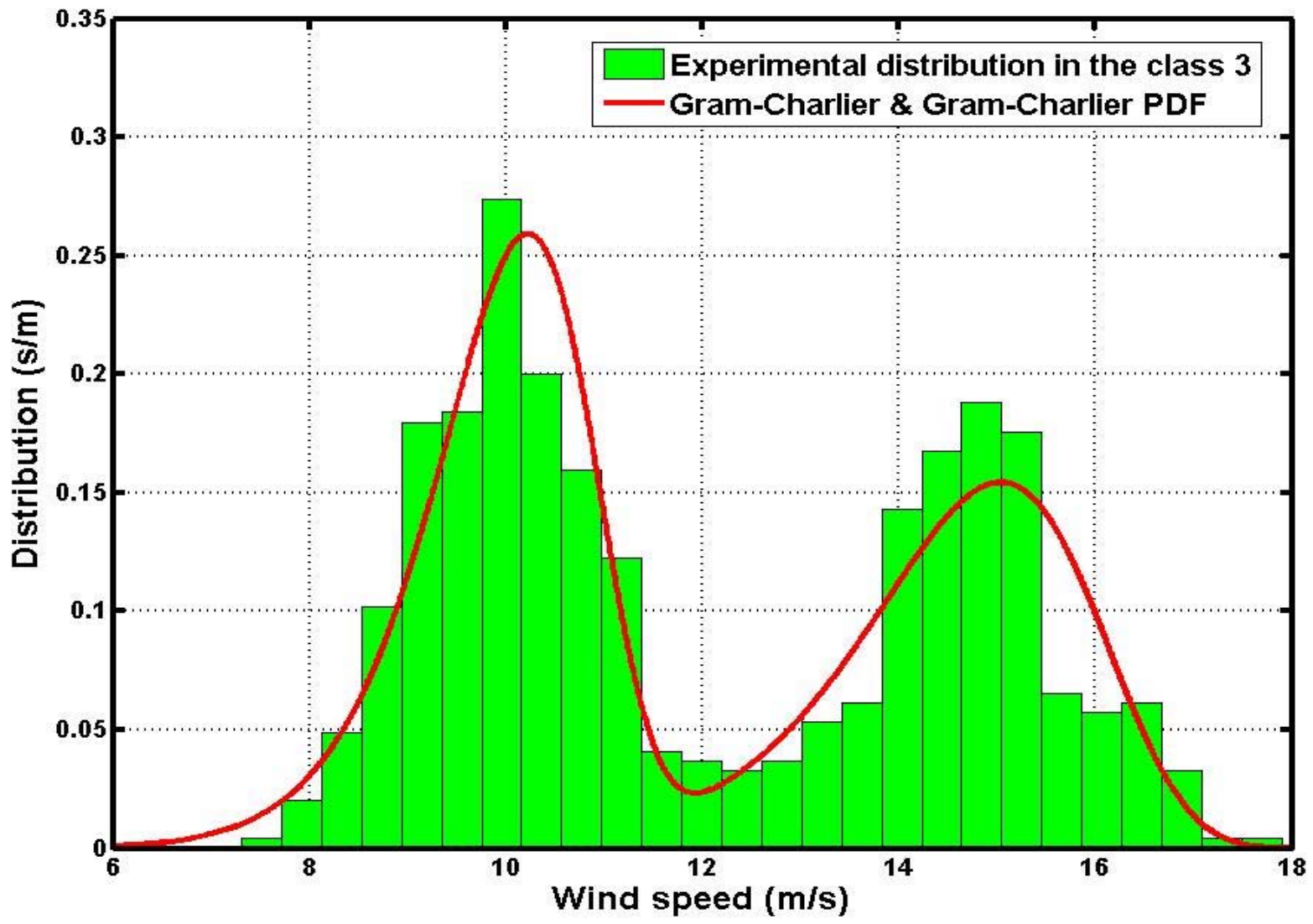

**Figure 12: Wind speed distribution in class 3.**

In the three cases, Kolmogorov-Smirnov test [12] was performed to compare the experimental wind speed distribution and the theoretical PDF.

## 6. Sequence of classes

1,000,118 sequences are classified into 3 classes; each measurement sequence can be replaced by its class number. We obtain a {1, 2, 3} - valued sequence of length 1,000,118. Figure 13 gives an example of sequence of holding time in each class: each number representing a class number. This can represents the wind regime evolution over the duration campaign.

It can be observed that this sequence has some interesting statistical properties such as an exponential residence time distribution in each class (figure 14) and also the transition from a class to another.

This leads us to think that such a sequence can be a path of a discrete Markov chain or a Hidden Markov Chain Model having 3 states *{1, 2, 3}*. This can be of interest for further research on wind energy prediction.

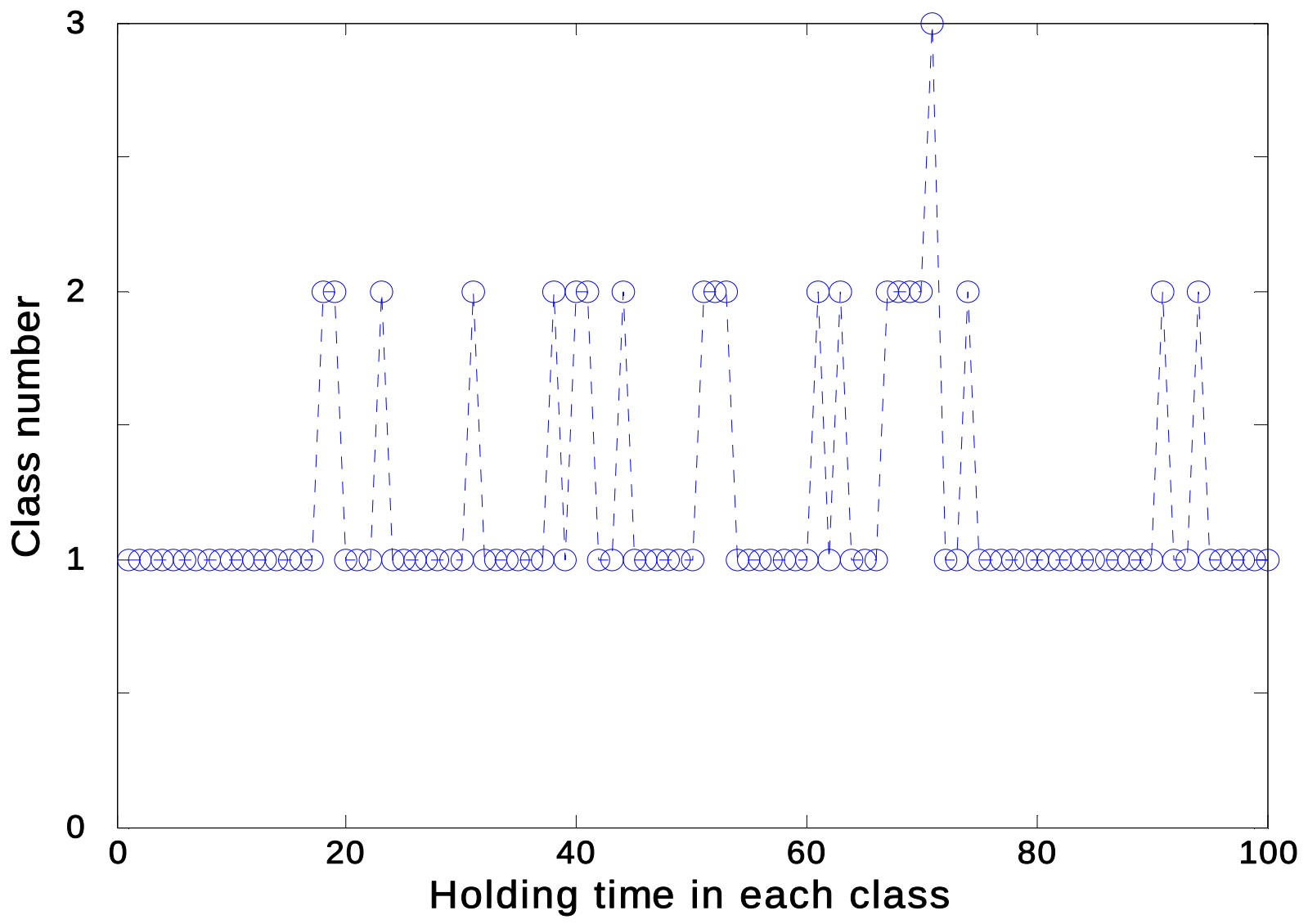

**Figure 13 : sequence of classes**

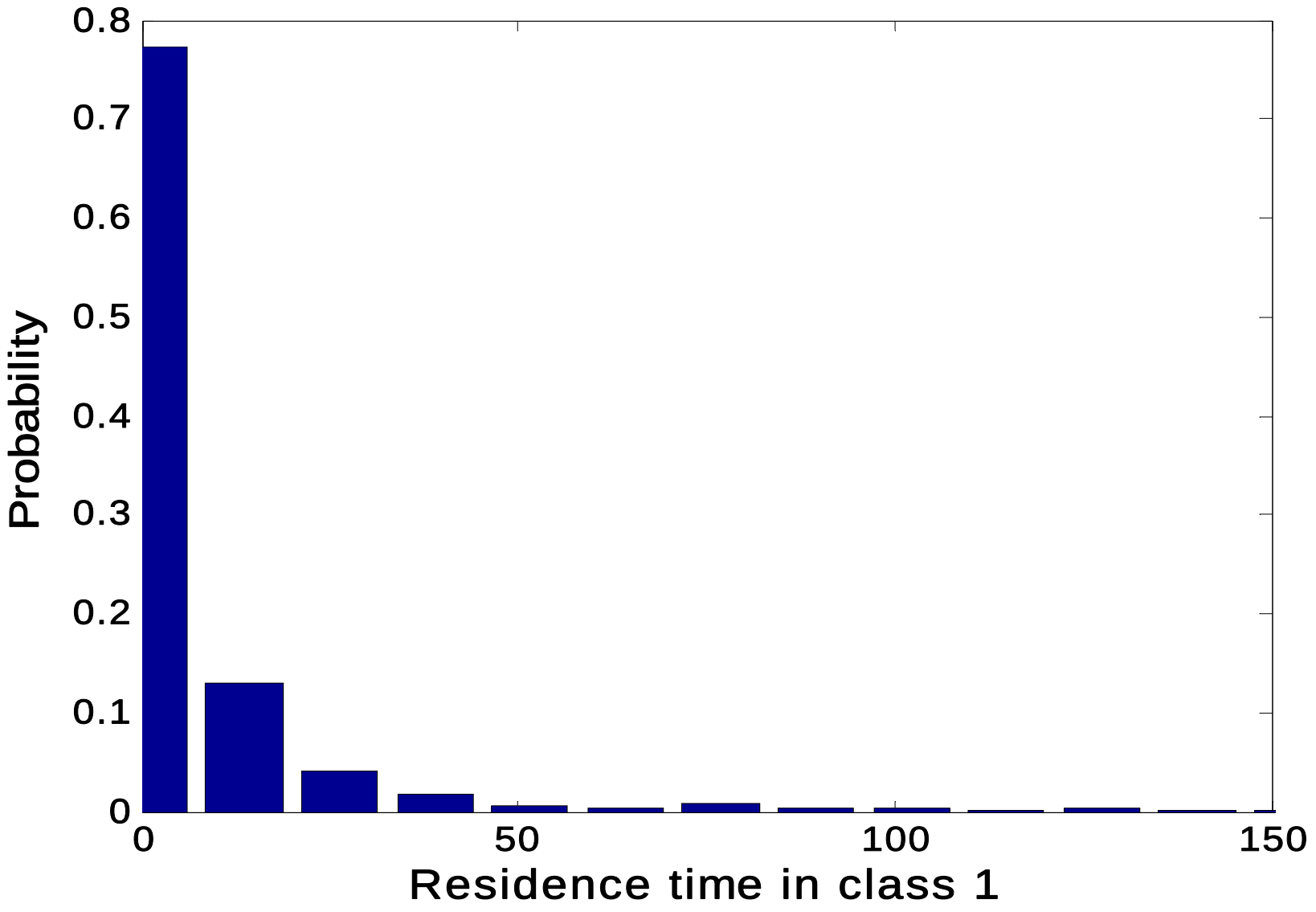

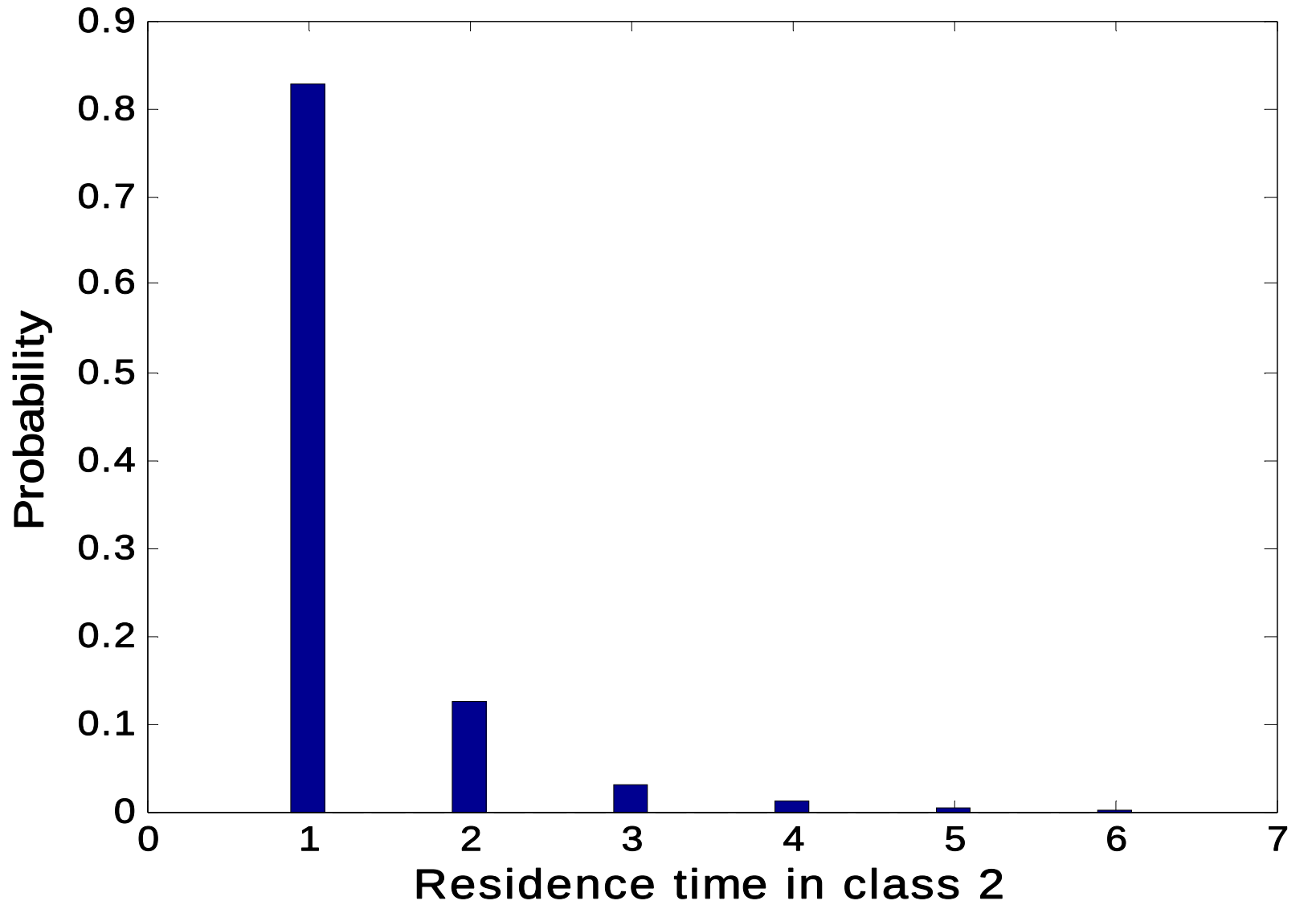

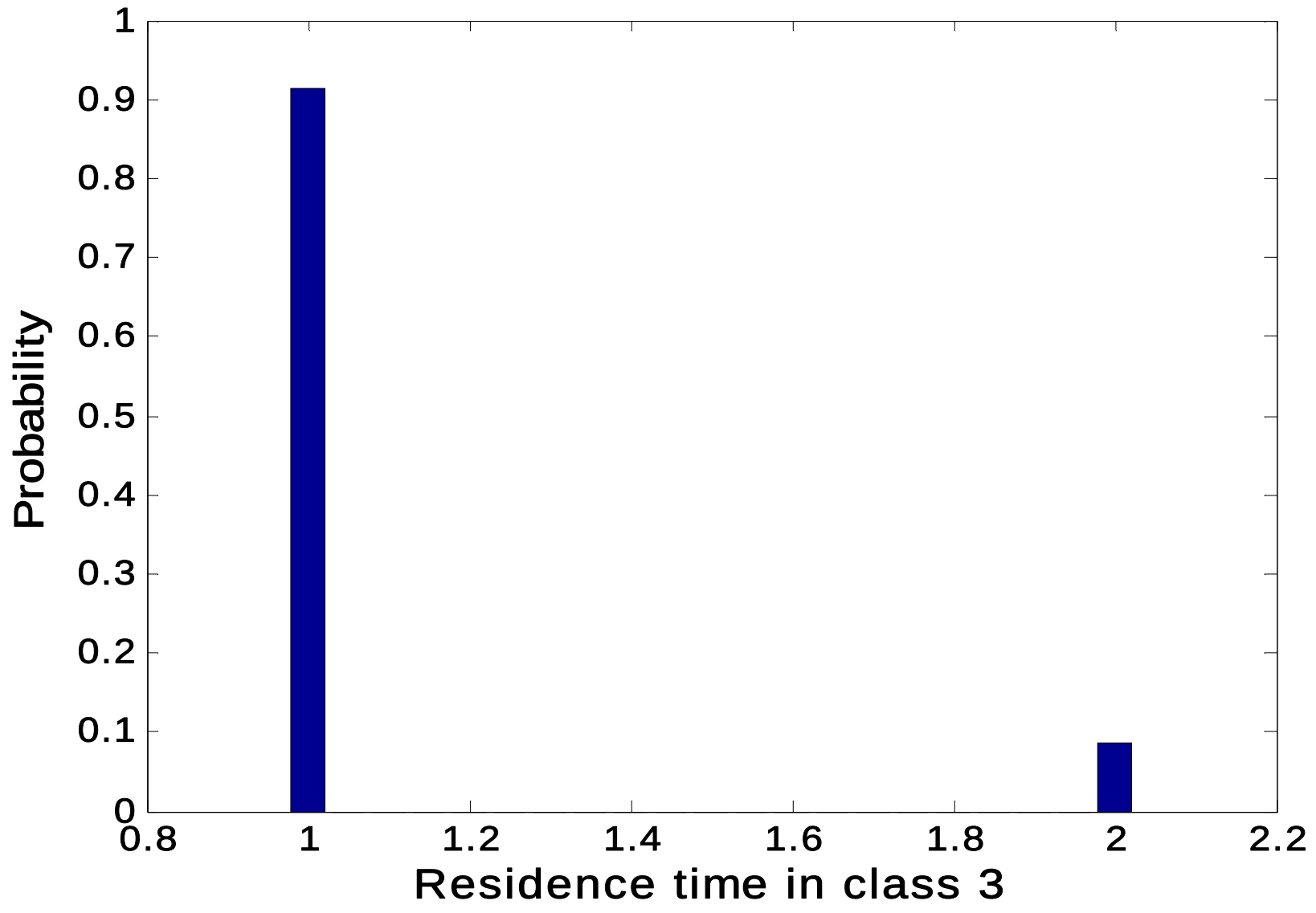

**Figure 14: Residence time in each class.**

## 7. Conclusions

This paper presents the results of wind speed PDF classification using Dirichlet distribution mixtures. We have first summarized. We have applied this new nonparametric method in order to classify the wind speed sequences in a statistical point of view. This enables us permitted to elaborate a forecasting tool for wind energy production on small times scales. We have used the Dirichlet distributions, one per class, to generate wind speed distributions as Dirichlet distribution are flexible enough to encompass a wide variety of distributional forms. The parameters of the mixture model are estimated using a variant of the Expectation Maximization algorithm, called the Stochastic Annealing Expectation Maximization. The method has been applied to wind speed measurements performed in Guadeloupe (16°2'N, 61°W) where important fluctuations can be observed even within a short period a few minutes. The results of the method have highlighted the existence of 3 classes of wind speed distribution:

1) A first class (90% of wind speed sequences) in which PDFs is symmetrical mono-modal modeled by a Gaussian PDF. The measurement sequence in this class, correspond to wind regime with a weak turbulent agitation.
2) A second class (9% of wind speed sequences) in which PDFs is dissymmetrical mono-modal PDF modeled by a Gram-Charlier function. A point of view meteorological, these times series are strong wind regimes.
3) A third class (1%of wind speed sequences) in which PDFs is bimodal PDF. This method is capable of drawing fine distinctions between the classes. Moreover, we have observed that mixtures of Gram Charlier type-A distributions fit to the wind speed distribution of each class. These measurement sequences can correspond to gust of wind or heavy shower. So they are drastically for wind energy production. Indeed these sequences can cause instabilities in case of intermediate power shortages.

The analysis of the sequence of classes leads us to think that the 10 minutes wind speed sequences are governed by a Hidden Markov Chain having 3 states *{1, 2, 3}*with some underlying unobservable *regimes* of wind speed sequence. This work presents some arguments for further investigations on wind speed forecasting, on time scales smaller than 1 hour.

## VII. References

[1] J. Bilmes. A gentle tutorial on the em algorithm including Gaussian and Baum-Welch. Technical report TR-97-021, 1997.

[2] R. Calif. Mesure et analyse de la vitesse du vent sur un site de production d'énergie éolienne en guadeloupe: Modèles pour la prévision sur des échelles de temps inférieures à l'heure. Phd thesis from Université des Antilles et de la Guyane, 2005.

[3] C. Celeux, D. Chauveau, and J. Diebolt. On stochastic version of the em algorithm. Technical report RR-2514 INRIA, 1995.

[4] R. Emilion. Classification et mélange de processus. Compte rendus Académies des Sciences Paris, 2002.

[5] M. D. Escobar. Estimating normal means with a dirichlet process prior. Journal of the American Statistical Association, 1994.

[6] M. D. Escobar and M. West. Bayesian density estimation and inference using mixtures. Journal of the merican Statistical Association, 1995.

[7] G. Giebel. On the benefits of the distributed generation of wind energy in europe. Phd thesis from the Carl von Ossietvsky Univertät Olenburg, 2001.

[8] G. Giebel, L. Landberg, G. Kariniotakis, and R. Brownsword. State of the art on methods and software tools for short term prediction of wind energy production. European Wind Energy Conference and Exhibition, 2003.

[9] Z. John and C. Chun-Ren. A study of the characteristic structures of strong wind. Atmospheric Research, 001.

[10] E. Jondeau and M. Rockinger. Gram-Charlier densities. Journal of Economics Dynamics and Control, 2001.

[11] K. K. Papagiannaki, N. Taft, and C. Diot. Impact of flows dynamics on traffic enginnering design principles. IEEE Infocom, 2004.

[12] F. J. Massey. The kolmogorov-smirnov test for goodness of fit. The American Statistical Association, 1956.

[13] P. Muller, A. Erkani, and M. West. Bayesian curve-fitting using multivariate normal mixtures. Biometrika, 1996.

[14] A. Oveissian, K. Salamatian, and A. Soule. Flow classification on short time scale. Technical report LIP6, 2003.

[15] S. Papoulis and S. Unnikrishna Pillai. Probability, random variables and stochastic processes. MC Graw Hill, 2002.

[16] S. Pavlos Georgilakis. Technical challenges associated with the integration of wind power into power systems. Renewable and Sustainable Energy Reviews, 2006.

[17] C. P. Robert. The bayesian choice. Springer, 2001.

[18] A. Soule, K. Salamatian, N. Taft, and K. Emilion, R. Papagiannaki. Flow classification by histograms or how to go on safari in the internet. Sigmetrics/Performance, 2004.

[19] D. Weisser and R. S. Garcia. Instantaneous wind energy penetration in isolated grids: concepts and reviews. Renewable Energy, 2004.

[20] O.A. Jamarillo et M.A. Borja. Wind speed analysis in la ventosa, mexico : a bimodal probability distribution case. *Renewable Energy*, 29 :1613{1630, 2004.